\documentclass[twocolumn]{aastex631}

\def\J1122{WISE\,J1122}

\bibpunct{(}{)}{;}{a}{}{,} 
\shorttitle{Main-oval aurorae in WISE\,J112254.72+255022.2}
\shortauthors{Guirado et al.}
\graphicspath{{./}{images/}}
\usepackage{appendix}

\begin{document} 

\title{Main-oval auroral emission from a T6 brown dwarf: observations, modeling, and astrometry }

\author[0000-0003-2722-1615]{J.C. Guirado}
\affiliation{Departament d'Astronomia i Astrof\'isica, Universitat de Val\`encia, C. Dr. Moliner 50, E-46100 Burjassot, Val\`encia, Spain}
\affiliation{Observatori Astron\`omic, Universitat de Val\`encia, Parc Cient\'ific, C. Catedr\'atico Jos\'e Beltr\'an 2, E-46980 Paterna, Val\`encia, Spain}

\author[0000-0002-5093-6208]{J.B. Climent}
\affiliation{Departament d'Astronomia i Astrof\'isica, Universitat de Val\`encia, C. Dr. Moliner 50, E-46100 Burjassot, Val\`encia, Spain}

\author[0009-0006-2594-3939]{J.D. Bergasa}
\affiliation{ Universidad Internacional de Valencia (VIU), C/ Pintor Sorolla 21, E-46002 Valencia, Spain}

\author[0000-0001-5654-0266]{M.A. Pérez-Torres}
\affiliation{Instituto de Astrof\'isica de Andaluc\'ia, Consejo Superior de Investigaciones Cient\'ificas (CSIC), Glorieta de la Astronom\'ia s/n, E-18008, Granada, Spain}
\affiliation{Facultad de Ciencias, Universidad de Zaragoza, Pedro Cerbuna 12, E-50009 Zaragoza, Spain }
\affiliation{School of Sciences, European University Cyprus, Diogenes street, Engomi, 1516 Nicosia, Cyprus}
        
\author[0000-0002-2736-9794]{J.M. Marcaide}
\affiliation{Real Academia de Ciencias Exactas, Físicas y Naturales de España, Calle Valverde 22, E-28004, Madrid, Spain } 

\author[0000-0001-6735-1655]{L. Pe\~{n}a-Mo\~{n}ino}
\affiliation{Instituto de Astrof\'isica de Andaluc\'ia, Consejo Superior de Investigaciones Cient\'ificas (CSIC), Glorieta de la Astronom\'ia s/n, E-18008, Granada, Spain} 

\begin{abstract}
 From a series of 5 GHz VLBA radio observations taken over 1-yr span, we present the detection of compact highly-polarized radio emission from the T6 brown dwarf WISE\,J112254.72+255022.2 compatible with electron cyclotron maser emission. Both the total and polarized lightcurves show variability in correspondence with a rotation period of 1.95$\pm$0.03\,hr. Comparison with models indicates that the quasi-steady radio emission of this brown dwarf is produced in circumpolar auroral rings, with remarkable similarity to the main-oval auroras in Jupiter. We have detected a large 100\% polarized flare in one of the VLBA epochs (2022.82) which may imply the existence of active longitudes in the auroral rings with a non-axisymmetric beaming cone radiation pattern, similar to the dusk/dawn asymmetries seen in the Jovian radio emissions. We also present a high-precision astrometric analysis of the sky motion of WISE\,J112254.72+255022.2, resulting in revised values of proper motion and parallax with an improvement in precision of one order of magnitude. The common kinematics of WISE\,J112254.72+255022.2 with its wide companion, the M-dwarf LHS\,302, is confirmed with submilliarcsecond precision, suggesting that this brown dwarf may have formed by gravitational fragmentation of the outer part of a protostellar disc around LHS\,302. The astrometric analysis imposes very tight bounds to the presence of low-mass companions around WISE\,J112254.72+255022.2, ruling out objects more massive than Saturn. Our results strengthen the analogy between radio emitting brown dwarfs and the magnetized planets of our solar system.
\end{abstract}


\section{Introduction}

Ultracool dwarfs are defined as cosmic bodies with spectral type later than M7V, encompassing very-low mass stars, brown dwarfs, and planetary objects \citep{1997AJ....113.1421K}. The M7V spectral type indicates the beginning of processes related with the decreasing effective temperature ($<$ 2700\,K), in particular the appearance of dust clouds, which significantly influence the atmospheric chemistry and its evolution (e.g. \citealt{2008MNRAS.391.1854H}). UCDs with the latest spectral types can be considered as scaled-up exoplanet analogs, excellent targets to better understand a wide range of phenomena related to formation mechanisms, low gravity, dusty atmospheres, or planetary magnetism \citep{2018Geosc...8..362C}.

UCDs are fully convective objects, with no shearing layer between convective and radiative
regions able to generate a magnetic field from the dynamo processes known to occur in stars with earlier spectral types. The discovery of radio emission from a M9V dwarf (\citealt{2001Natur.410..338B}) constituted a relative surprise, as the radio detection violated the empirical G\"udel-Benz relationship between X-rays and radio luminosities, in turn suggesting that magnetic activity and, correspondingly, radio emission were not precluded in UCDs. Subsequent radio surveys showed 
that $\sim$10\% of the observed UCDs are radio emitters
(\citealt{2015ApJ...808..189W}; \citealt{2016ApJ...830...85R}; \citealt{2024MNRAS.527.6835K}). This radio emission is consistent with the presence of a quiescent component, attributed to unpolarized gyrosynchrotron emission, and another pulsed, highly polarized emission produced by the electron cyclotron maser instability
(ECM; \citealt{1982ApJ...259..844M}; henceforth cited as MD82). Highlighting the analogy of some UCDs with (exo)planets, the ECM mechanism is known to operate in the magnetized planets of the solar system (e.g., \citealt{2011JGRA..116.4212L}; \citealt{2017A&A...604A..17M}), and it is postulated to be present in exoplanets (\citealt{2007P&SS...55..598Z}; \citealt{2011MNRAS.414.2125N}).

Radio emission of UCDs is frequently associated with short rotation periods \citep{2017ApJ...846...75P}. However, not all rapid rotator UCDs are detectable radio emitters \citep{2012ApJ...746...23M}, emphasizing the importance of other factors at play, such as the plasma conditions and the strength and topology of the magnetic field. In principle, the association between low multipole order fields and radio emission is supported by both the dipolar magnetic fields seen in the planets of the solar system, and the success in modeling the emission of radio luminous UCDs assuming a dipolar configuration (i.e., \citealt{2017MNRAS.470.4274T}). However, global non-axisymmetric magnetic topologies have been inferred \citep{2015ApJ...802..106L}, suggesting that more complex magnetic configurations might also be present in radio emitting UCDs.

The enhanced sensitivity of interferometers operating primarily at GHz-frequencies (J-VLA, ATCA), and the advent of SKA precursors (LOFAR, ASKAP), optimized at MHz to GHz frequencies, have boosted the detection of UCDs, showing evidence of radio emission in objects with increasingly cooler spectral types including, among others, a T2.5-type candidate to planetary mass object \citep{2018ApJS..237...25K}, a T5.5\,+\,T7.0 binary \citep{2023A&A...675L...6V}, and a T8-dwarf \citep{2023ApJ...951L..43R}. 

\begin{table*}[!t]
    \caption{Journal of observations} 
    \label{table:obs} 
    \centering 
    \begin{tabular}{l c c c c } 
    \hline\hline 
        Antennas & Project / Segment & Observing Date & UT range & Synthesized beam  \\ 
                 &         &            &       & (mas,$^{\circ}$) \\ 
    \hline
        VLBA & BG278\,A  & 08 Apr. 2022 (2022.27) & 03:30-07:30 & $4.1\times1.8$, --0.9 \\
        VLBA$^a$ & BG278\,B  & 06 Jul. 2022 (2022.51)& 21:30-01:30 & $3.8\times1.6$, --2.6 \\
        VLBA$^a$ & BG278\,C  & 25 Oct. 2022 (2022.82)& 14:30-18:30 & $3.8\times1.5$, --3.3 \\
        VLBA$^a$ & BG278\,D  & 09 Jan. 2023(2023.02) & 09:30-13:30 & $3.6\times1.5$, --3.0 \\
        VLBA$^b$ & BG278\,E  & 15 Apr. 2023 (2023.29) & 03:00-07:00 & $4.2\times2.2$, --7.9 \\

    \hline 
    \end{tabular}
    \tablecomments{
    $^{\mathrm{a}}$ VLBA-KP did not participate.$^{\mathrm{b}}$ VLBA-MK did not participate.\\}
    \end{table*}

In addition to the above-mentioned radio interferometers, the high brightness temperature of the radio emission measured in a number of UCDs (see references below) enables the use of the very-long-baseline interferometry (VLBI) technique. The high spatial resolution that VLBI provides is a suitable tool to constrain both the magnetic and kinematic properties of the UCDs; actually, VLBI observations have contributed to remarkable results, namely, the establishment of direct limits on the radio emission brightness temperature (TVLM\,513$-$46546; \citealt{2009ApJ...706L.205F}), the measurement of precise dynamical mass in a binary system 
(2MASS\,J0746+2000AB; \citealt{2020ApJ...897...11Z}), the discovery of the first planetary companion from radio astrometry 
(TVLM\,513$-$46546; \citealt{2020AJ....160...97C}), and the discovery of the first extrasolar radiation belts (LSR\,J1835+3259; \citealt{2023Natur.619..272K}, \citealt{2023Sci...381.1120C}). 

In this context, the application of VLBI astrometry to the coolest UCDs results of special interest. T-dwarfs are excellent targets for astrometric searches of exoplanets, since their low mass facilitate the presence of lower-mass planets (Saturn to Earth-mass) orbiting them;
additionally, the sub-milliarcsecond precision inherent to VLBI astrometry allows to detect companions throughout a wide range of planetary masses, covering objects formed through different mechanisms, i.e., from giant planets (mass ratio $q$\,$\sim$\,0.1), probably formed as binary systems via fragmentation of cloud cores 
\citep{2004ApJ...617..559P}, to Saturn-mass planets ($q$\,$\sim$\,0.01), probably formed as {\it bona fide} planets via a protoplanetary disk \citep{2012ApJ...755...67H}.

Given the considerations above, and to increase the statistics of UCDs with compact radio emission, we initiated a VLBI program to observe nearby, fast-rotating ultracool dwarfs, covering spectral types from M7 to T6. In this paper, we focus on WISE\,J112254.72+255022.2 
(\citealt{2011ApJS..197...19K}; hereafter \J1122), a T6-type brown dwarf located at 15.9 pc \citep{2020AJ....159..257B}. The radio emission of \J1122 was discovered at GHz-frequencies with the Arecibo telescope (\citealt{2016ApJ...821L..21R}; henceforth cited as RW16), with follow up detection with the VLA (\citealt{2017ApJ...834..117W}; henceforth cited as WGB17). Periodic radio bursts from \J1122 have been observed with both instruments, although reporting different rotation periods: 17.3 minutes, from pulsar timing techniques using Arecibo, and 116 minutes, from the  
lightcurve measured with the VLA. In any case, the extreme rotation rate of \J1122 along with its late spectral type convert this object in a prime target to investigate magnetic phenomena in the realm of brown dwarfs.

We report on the detection of milliarcsecond-scale highly-polarized radio emission from \J1122 with variability in correspondence with its rotation period. This radio emission is compatible with electron cyclotron maser (ECM) emission produced in circumpolar auroral rings, with remarkable similarity to the main-oval auroras in Jupiter. We also present a high-precision astrometric analysis of the sky motion of \J1122, resulting in revised values of proper motion and parallax. Bounds to the presence of lower-mass companions are discussed.


\section{Observations and data reduction}\label{sect:observations_datareduciton}

We carried out a series of Very Long Baseline Array (VLBA) observations at C-band (5 GHz) during 2022 and 2023 (project code BG278, see Table \ref{table:obs} for details). We used a standard continuum setup for
the 6 cm band, with eight sub-bands of width 128 MHz recorded
in dual circular polarization at 2-bit per sample for a total data
rate of 4 Gbps per antenna. We used the phase-reference technique, interleaving scans of \J1122 and the extragalactic source J112553.7$+$261019 (angular separation of 0.75$^{\circ}$), the latter with coordinates included in the 3rd realization of the International Celestial Reference Frame (ICRF3; \citealt{2020A&A...644A.159C}). The duty cycle lasted 6\,min, with 4\,min on the target, and approximately 1\,min on the phase calibrator, providing a total integration time on \J1122 of $\sim$220\,min per epoch.

\begin{figure*}
\centering
   \includegraphics[trim={-1.0cm 0cm 0.3cm 0cm},clip, width=0.99\linewidth]{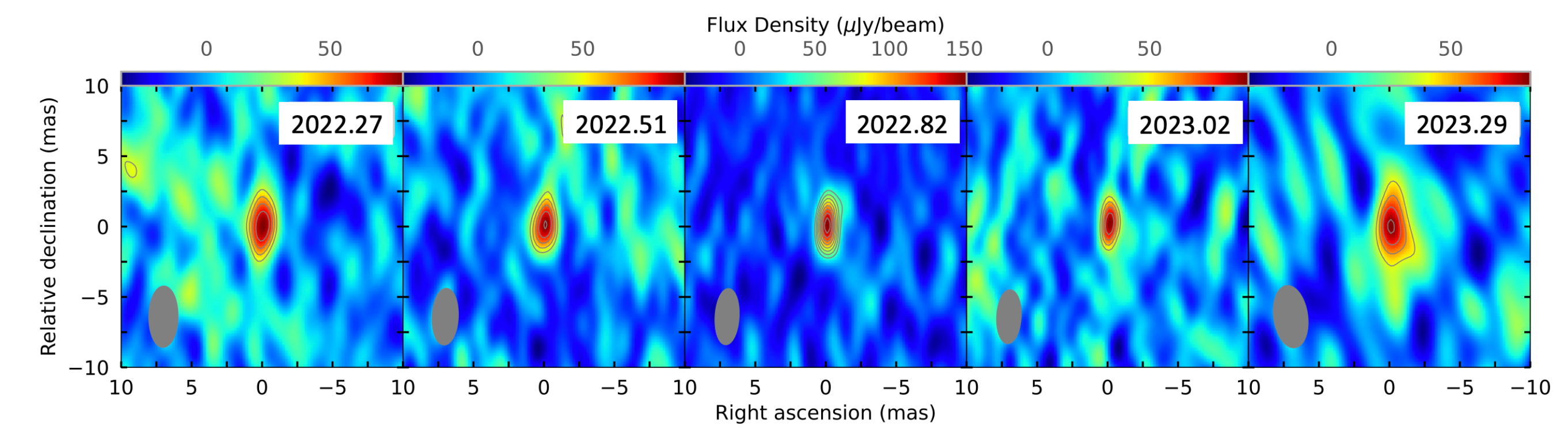}
 \caption{Reconstructed total flux (Stokes I) images centered on \J1122 coordinates corresponding to the each of the VLBA observing epochs. The contours represent detection levels at 3$\sigma$, 4$\sigma$, 5$\sigma$, 6$\sigma$, etc. The grey ellipses in the lower-left corners represent the FWHM beam sizes (image properties are summarized in Table 2).}
    \label{fig:imaps}
\end{figure*}

\begin{figure*}

\centering
   \includegraphics[trim={0cm 0cm 0cm 0cm},clip, width=1.0\linewidth]{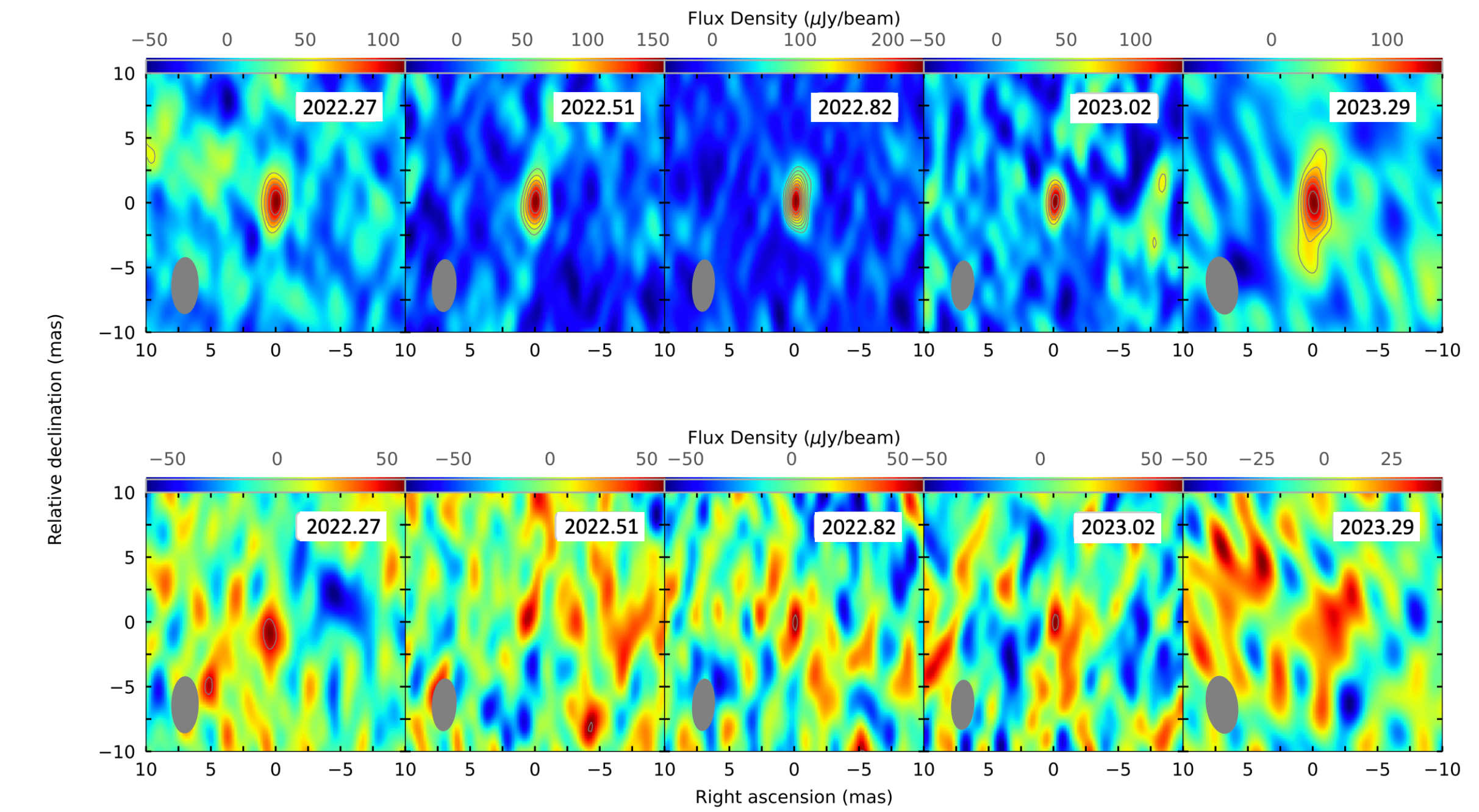}

   \caption{Reconstructed RCP (upper row) and LCP (bottom row, showing non-detections) images centered on the coordinates of \J1122 corresponding to the each of the VLBA observing epochs. The contours represent detection levels at 3$\sigma$, 4$\sigma$, 5$\sigma$, 6$\sigma$, etc. The grey ellipses in the lower-left corners represent the FWHM beam sizes (image properties are summarized in Table 2).} 
    \label{fig:rmaps}
\end{figure*}

\begin{table*}[t]
\centering
\caption{\J1122 image parameters} 
\label{table:flux} 
    \centering 
    \begin{tabular}{l c c c c c c c } 
    \hline\hline 
 & \multicolumn{2}{c}{Stokes I} 
& \multicolumn{2}{c}{RCP}
& \multicolumn{2}{c}{LCP} \\

Project & S$^{(a)}_{\rm{peak}}$ & $\sigma^{(b)}$ & S$^{(a)}_{\rm{peak}}$  & $\sigma^{(b)}$ & S$^{(a)}_{\rm{peak}}$ & $\sigma^{(b)}$ \\
& [$\mu$Jy/beam] & [$\mu$Jy/beam] & [$\mu$Jy/beam] & [$\mu$Jy/beam] & [$\mu$Jy/beam] & [$\mu$Jy/beam]\\
\hline
BG278A  & 79$\pm$15 & 12 & 114$\pm$18 & 12 &66 & 15 \\
BG278B  & 99$\pm$12 & 14 & 157$\pm$18 & 14 &$<57^{(c)}$ & 19 \\
BG278C  & 152$\pm$18 & 14 & 241$\pm$25 & 14 &62& 16  \\
BG278D  & 98$\pm$14 & 14 & 134$\pm$17 & 14 &60 & 16 \\
BG278E  & 93$\pm$20 & 14 & 148$\pm$20 & 14 &$<57^{(c)}$ & 19 \\
    \hline 
    \end{tabular}
    \tablecomments{
    $^{(a)}$ Peak of brightness. $^{(b)}$ Root mean squared (rms) noise. $^{(c)}$ 3\,$\sigma$ upper bound.}
    \end{table*}
    
The data were calibrated using the NRAO 
Astronomical Image Processing System (\texttt{AIPS}; \citealt{1996ASPC..101...37V}) following routines included in the \texttt{VLBARUN} pipeline:
(i) we performed amplitude calibration using system
temperatures and antenna gains provided by each station; (ii) we corrected the ionospheric
dispersion using GPS-based Global Ionospheric Maps; (iii) we updated the 
Earth orientation parameters at the time of the observations; (iv) 
we corrected for the parallactic angle; (v) we performed a fringe search on the calibrator integrated over each scan to remove residual contributions to the
phases, and (vi) we interpolated these solutions from the calibrator onto the target data.

The calibrated visibilities were subsequently imaged and deconvolved with the Common Astronomy Software Applications (CASA) package \citep{2007ASPC..376..127M}. We employed naturally-weighted, wide-field images to search for compact radio emission of the brown dwarf \J1122, by examining the area around its anticipated coordinates. The latter were obtained from the (radio) position given in WGB17 propagated to each observing epoch using the proper motion and parallax values provided by 
\citet{2011ApJS..197...19K}. In all five VLBA epochs, a single unresolved point source is detected well within the uncertainties of the expected positions of \J1122 (initially $\sim$0.3" precise, but substantially improved after subsequent VLBA detections); considering the large proper motion of this source (exceeding 1"/yr), the association of our VLBA detections with radio emission from \J1122 is corroborated. Finally, we used the \texttt{tclean} algorithm to obtain phase-referenced, channel-averaged, total flux images of \J1122 (see Fig. \ref{fig:imaps}). We again performed the procedure above to obtain images at both right and left circular polarization (RCP and LCP, respectively; see Fig. \ref{fig:rmaps}). Image parameters are shown in Table \ref{table:flux} . We emphasize that, during this phase-reference mapping
process, the positional information of \J1122 with respect to
the external quasar is conserved, thus relating the position of
\J1122 to the ICRF (see Sect. 4.4).

\section{Results}\label{sect:results}

\begin{figure}[!htb]
    \centering
        \includegraphics[trim={3cm 0cm 3cm 0cm},clip, width=1.0\linewidth]{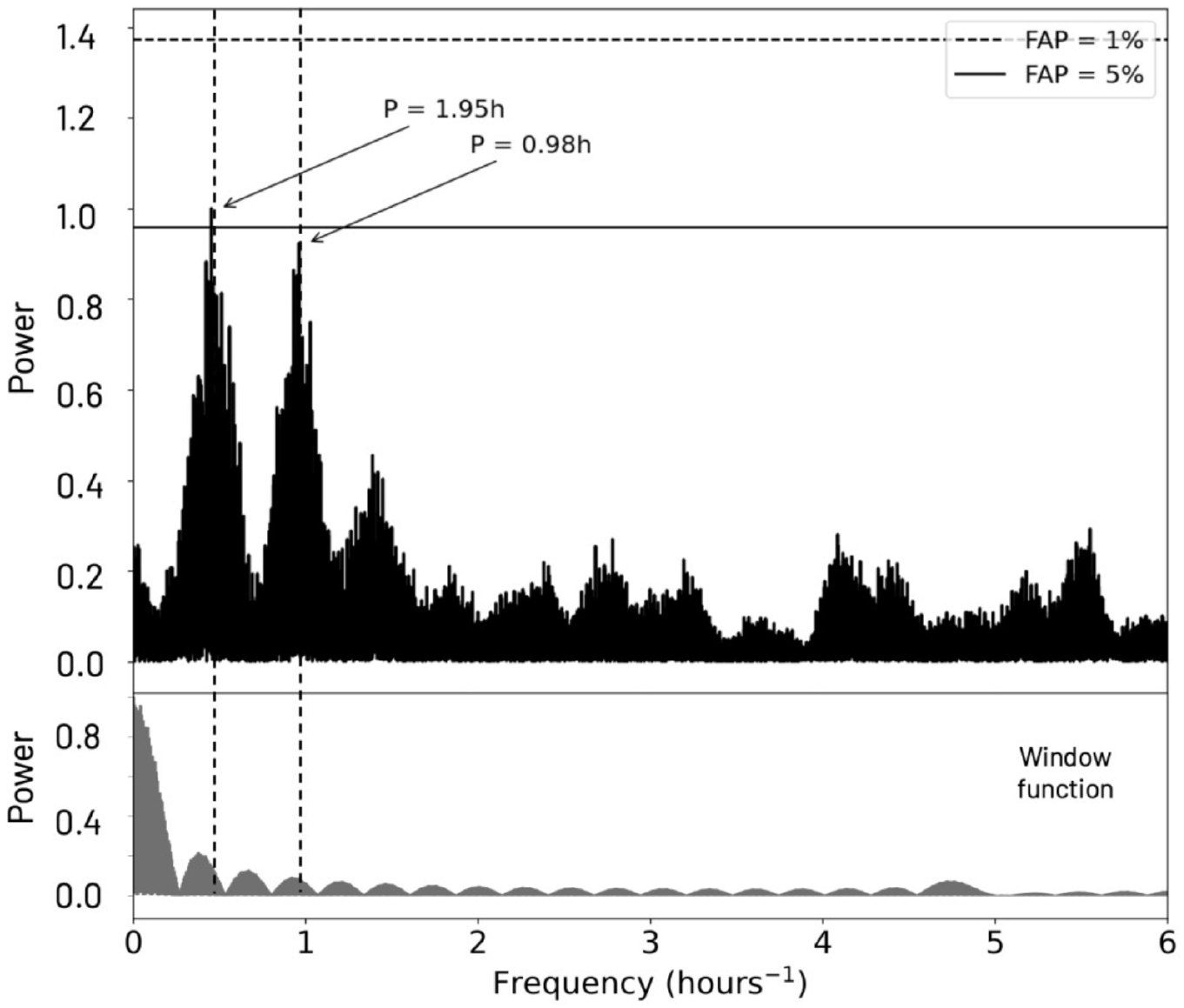}
    \caption{
GLS periodogram of the spectrally-averaged time series of RCP data sampled with a time resolution 
of $\sim$180\,s. Time series of the five observing epochs have been combined in the periodogram. False alarm probabilities (FAP) of 1\% and 5\% are plotted for reference. We interpret the largest peak (1.95\,hr) as the most probable rotation period of \J1122 (in coincidence with WGB17's estimate) while the second peak (0.98\,hr) would show the presence of intra-period variability. The periodogram of the window function is also shown.}
\label{fig:gls}
\end{figure}

Fig. \ref{fig:imaps} reveals detections in Stokes I of an unresolved source in all observing epochs. The flux density averaged $\sim$80\,$\mu$Jy, except for segment C, which shows a significant increase in flux reaching 137$\pm$18\,$\mu$Jy. These are the first VLBI images of \J1122, which confirm the presence of compact and persistent radio emission in such a cool T6 object. 
Circularly polarized maps (Fig. \ref{fig:rmaps}) show that RCP emission dominates in all BG278 segments, 
with flux densities ranging from 100 to 220\,$\mu$Jy, again with segment C showing the maximum values. In contrast, LCP emission is only marginally detected in segments A, C, and D with flux densities slightly above the detection threshold ($3\sigma$$\sim$\,50\,$\mu$Jy/beam). This points towards highly polarized radio emission.

\begin{figure*}[!htb]
    \centering
\includegraphics[trim={4cm 0cm 4cm 0.5cm},clip, width=0.9\linewidth]{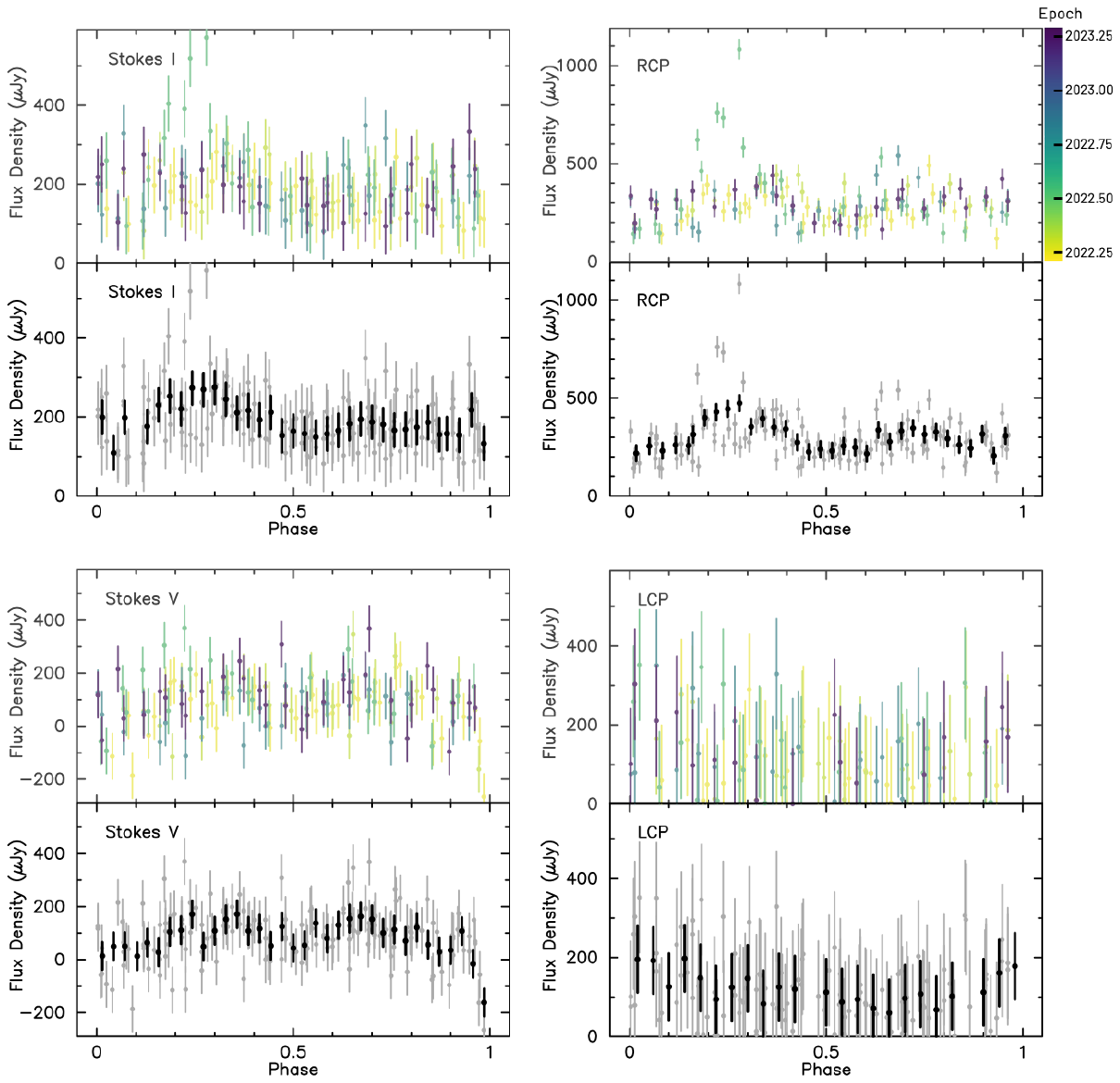}
    \caption{Periodicity of the flux density of \J1122. Stokes I/V and RCP/LCP lightcurves of all epochs phase wrapped to a rotation period of 1.95\,hr. Stokes V and LCP lightcurves are obtained from the combination of Stokes I and RCP observed data. For each panel, the upper plots show the lightcurve of each of the five observing epoch according to the color scale; the lower plots show the averaged binned values (black) along with the individual curves (grey). 
    An intra-period sinusoidal trend is present in the data (reflected in the secondary peak of the periodogram in Fig. \ref{fig:gls}); this sinusoid is most evident in the RCP lightcurve, peaking in rotation phases $\sim$0.3 and $\sim$0.7 with 6$\sigma$ and 3$\sigma$ over the rms noise, respectively.}
\label{fig:lcurves}
\end{figure*}

Indeed, previous observations already reported circularly polarized pulsed radio emission from \J1122. 
WGB17 detected 4--8\,GHz time-variable radio emission from \J1122 showing primarily RCP bursts ($\sim$500\,$\mu$Jy) alongside a weaker LCP emission ($\sim$100\,$\mu$Jy). These authors report a periodicity of $\sim$116\,min. On the other hand, in the course of a 5\,GHz search from flaring emission in L and T dwarfs conducted with the Arecibo telescope, \citet{2016ApJ...821L..21R} detected several 15\%-100\% LCP bursts from \J1122 with flux densities $\sim$1.5--3\,mJy, and an estimated period of 
$\sim$17.3\,min, a value near the rotational breakup limit, assuming \J1122 to be a highly-oblate object older than 1\,GHz (RW16).
In turn, we examined the presence of radio bursts and time variability in our VLBA flux density data.

\subsection{Time variability}

We used the CASA routine \texttt{uvmodelfit} to obtain time series of spectrally-averaged Stokes I and RCP flux values of \J1122 at all epochs. We selected the time resolution given by each scan length ($\sim$180\,s), as a balance between SNR and sensible monitoring. Given the low SNR of the LCP data, \texttt{uvmodelfit} did not produce usable Stokes V and LCP time series. Instead, for the sake of a better representation of the polarization properties of \J1122, we recovered both time series using the well-known relationships in the quasi-monochromatic approach $I=\frac{1}{2}(RCP+LCP)$, and $V=\frac{1}{2}(RCP-LCP)$. 

Given the previous lightcurves reported in RW16 and WGB17, the radio emission of \J1122 is expected to be rotationally modulated; therefore, we searched for periodicities in our flux density data to appropriately combine the five observing epochs folded to the most favorable rotation rate. We used of a Lomb-Scargle approach 
(\citealt{1976Ap&SS..39..447L}; \citealt{1982ApJ...263..835S}) to find the period(s) inherent to the time evolution of the radio flux density. We present in Fig. \ref{fig:gls} the resulting periodogram for the RCP data set, that with the highest SNR,  
where we found two dominant peaks with a false alarm probability $\sim$1\,\%: the first one corresponds to $1.95\pm0.03$\,hr and the second one at $0.98\pm0.04$\,hr (i.e. 117 and 59\,min, respectively); the uncertainties were derived from the full width half maximum of the peaks and the average SNR of the data points \citep{2018ApJS..236...16V}. Our longer period of 1.95\,hr is in excellent agreement with the WGB17 estimate from VLA observations, and we interpret it as the putative rotation period of \J1122, while the shorter period of 0.98\,hr is associated to intra-period flux variations, as discussed in Sect. 3.2. Our data do not favor the extremely short period of 17.3\,min reported by RW16; these authors also proposed an alternative rotation rate of 0.863\,hr, which is close, but not coincident (3\,$\sigma$ difference), to the second peak of our periodogram ($0.98\pm0.04$\,hr).

\subsection{Total and polarized lightcurves }

We present our VLBA lightcurves in Fig. \ref{fig:lcurves}, where the data of all observing epochs are phase wrapped to a rotation period of 1.95\,hr. Binned median values are also shown in Fig. \ref{fig:lcurves}; we understand that this averaging may smooth out rapid flares and/or underestimate the flux values at certain rotation phases, but it highlights longer-term trends, providing a valuable view of the quasi-steady emission of \J1122 averaged over one year, the time span of our observations. 

The averaged RCP lightcurve shows a sinusoidal trend that peaks in rotation phases $\sim$0.3 and 0.7. Since this is the predominant polarization handedness in the radio emission of \J1122, both peaks are also visible in Stokes I and V; actually, Stokes V is positive (i.e., preference for RCP) during 90\% of the rotation period. The shorter period of 0.98\,hr in our periodogram would reflect the time separation of these two maxima within the same rotation. In contrast, the reconstructed lightcurve in LCP shows almost negligible flux values for most of the period, showing a weak increase in coincidence with the depths of RCP around rotation phases 0.1 and 0.9. The presence of these sinusoidal patterns in the averaged flux densities suggests that whatever the acting radiation mechanism, it must be capable of producing polarized radio emission in most of the rotation phases, with stable intensity on average, on time scales of one year. This statement is supported by the similarity of the RCP/LCP VLBA lightcurves with those obtained with the VLA in 2016 (Fig. 1 in WGB17) which also present a RCP double-peaked profile; being the similarity of both lightcurves a remarkable coincidence, we notice that both the uncertainties associated with the rotation period and the long time baseline between our data and WGB17's do not allow us to phase-align both profiles.  

\begin{figure}[!ht]
    \centering
     \includegraphics[trim={0.5cm 2cm 10cm 17cm},clip, width=1.0\linewidth]{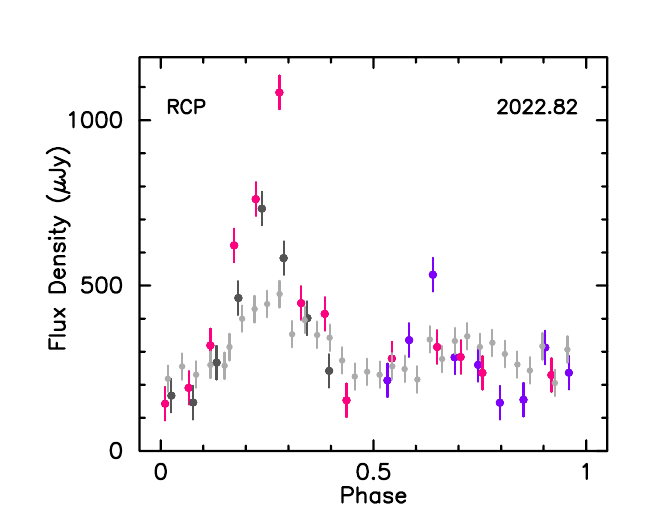}
    \caption{Detail of the RCP lightcurve corresponding to segment C (epoch 2022.82). Different colors correspond to different, but consecutive periods (chronologically, purple -- red -- dark grey). The averaged lightcurve is shown in light grey.}
\label{fig:Cflare}
\end{figure}

Having said the above, we note that the VLBI lightcurves present excursions from the average values as a consequence of the variable nature of the radio emission; the most remarkable feature is the 100\% RCP burst with flux density $>$1\,mJy seen in phase 0.3 in segment C (see Fig. \ref{fig:Cflare}, where, for clarity, the lightcurve for this particular epoch 
is plotted separately). We discuss this particular feature in Sect. 4.3.

\section{Discussion}

\subsection{The radio emission of \J1122}

The average flux densities stated in the previous section imply a mean radio
luminosity of $\sim$10$^{13.7}$\,erg\,s\,$^{-1}$\,Hz$^{-1}$ at 15.9\,pc. Taking the 
flux density measured in the large burst in segment C, the radio luminosity raises 
to $\sim$10$^{14.5}$\,erg\,s\,$^{-1}$\,Hz$^{-1}$, similar values to those reported in 
RW16 an WGB17, and comparable to other radio emitting T-dwarfs 
(\citealt{2023ApJ...951L..43R}; \citealt{2023A&A...675L...6V}). 

If the emitting region is unresolved, the resolution of an interferometer and the measured flux density provide a lower limit of the brightness temperature of the radio emission \citep{2015A&A...574A..84L}. For the case of \J1122, based solely on these interferometric arguments, our data impose that $T_B > 2.5 \times 10^{8}$\,K. Further assuming that the size of the emitting region corresponds to the typical size of a T8 object ($\sim$1\,R$_{\rm{J}}$; \citealt{2002A&A...382..563B}), the lower bound of the brightness temperature becomes $T_B > 8 \times 10^{10}$\,K, comparable to the more stringent bound of $T_B > 4 \times 10^{11}$\,K determined by RW16 based on the intense flares detected with Arecibo. 

The smoothly varying behaviour of the lightcurves in Fig. \ref{fig:lcurves} may lead to consider gyrosynchrotron emission as a possible radiation mechanism. However, gyrosynchrotron emission, canonically associated with a quiescent, unpolarized component of the UCD radio emission, is unlikely to account for both the high $T_B$ above and the persistent circular polarization of \J1122. Rather, the explanation of these properties requires a coherent mechanism that powers the radio emission. The two processes typically considered are plasma emission and ECM (MD82).

Plasma radiation originates from plasma waves excited by high-energy electron density (Langmuir) waves \citep{1969SvPhU..12...42K}; plasma emission is radiated at the harmonics of the plasma 
frequency $\nu_p$ defined as
$\nu_p=9000\,n^{1/2}$\,Hz, with $n$ the plasma density in cm$^{-3}$. The brightness temperature is above 10$^{13}$\,K for a wide range of plasma conditions; emission at the first harmonic could be 100\,\% circularly polarized (e.g., \citealt{2001A&A...374.1072S}). ECM emission is produced from a population of high-energy electrons with a loss-cone anisotropy (i.e. deficiency of electrons with small pitch angle); the maser emission is activated at the first harmonics of the gyrofrequency $\nu_g=2.8\times10^6\,B$\,MHz (with $B$ the magnetic field in Gauss) provided 
that $\nu_g\gg \nu_p$ (MD82). The maser growth rate strongly depends on the angle of the wave vector $\hat{k}$ with the magnetic field, therefore producing highly beamed radiation that is emitted following a hollow-cone pattern attached to the local magnetic field line. ECM radiation is strongly circularly polarized reaching extremely large brightness temperatures of $T_B \sim 10^{18}$\,K (MD82). The handedness of the polarization depends of the 
magneto-ionic mode amplified by the ECM mechanism (either ordinary, $o$-mode, or 
extraordinary, $x$-mode, in turn depending of the direction of the local magnetic field). Our data do not allow us to distinguish which magneto-ionic mode is favored (different plasma conditions suppress one mode or the other). However, regardless of the mode being excited, radiation originating in opposite magnetic hemispheres will have opposite handedness. 

Therefore, both coherent mechanisms justify our high $T_B$ limits and high circular polarization properties. However, the rotational modulation seen in our averaged lightcurves (Fig. \ref{fig:lcurves}) is naturally explained by the beamed ECM emission instead, whose direction is determined by the magnetic field lines. Effectively, the modulation is produced by the changing relative geometry between the line of sight and the magnetic field line(s) as the brown dwarf rotates. Such a directivity is difficult to justify with the non-beamed plasma emission. Thus, although the presence of plasma radiation cannot be ruled out, our data favor the ECM mechanism as the main source of compact radio emission in \J1122. Assuming that ECM is radiated at the first harmonic of 
$\nu_g$, and taking our largest observing frequency band (5.1\,GHz), we obtain a lower bound for the magnetic field of $B > 1.8$\,kG (similar to the values reported in RW16 and WGB17). 
Likewise, using the condition for ECM $\nu_g\gg\nu_p$, we find that the electron density in the region where the maser 
is generated should accomplish that $n \ll 3\times 10^{11}\,\rm{cm}^{-3}$; this condition is likely to be met in \J1122 considering the plasma densities estimated for late-type M-dwarfs ($\sim10^{5-7}\,\rm{cm}^{-3}$; \citealt{2022A&A...660A..65C}) or even those of massive hot magnetic stars 
($\sim10^{9}\,\rm{cm}^{-3}$; \citealt{2020MNRAS.499L..72L}).

\begin{table*}[t]
\caption{Inferred geometric parameters of the auroral ring model for \J1122}
\label{table:mcmcresults} 
    \centering 
    \begin{tabular}{lll} 
    \hline\hline 
    Parameter & Prior & Posterior \\ \hline
Spin axis inclination ($i$): & $\mathcal{U}(0,90)$  & $86.8^{\rm{o}}$  $\pm$  $0.6^{\rm{o}}$ \\
Magnetic axis obliquity ($\beta$):& $\mathcal{U}(0,60)$  & \,\,\,$2.6^{\rm{o}} $  $\pm$  $ 0.5^{\rm{o}}$ \\
Magnetic colatitude of the auroral ring ($\theta_B$): & $\mathcal{U}(0,40)$& \,\,\,$2.7^{\rm{o}} $  $\pm$  $ 0.6^{\rm{o}}$ \\
Hollow cone half-opening angle ($\alpha$): &$\mathcal{U}(30,90)$ & $86.0^{\rm{o}} $  $\pm$  $ 0.7^{\rm{o}}$ \\
Hollow cone thickness ($\Delta\alpha$): & $\mathcal{U}(0,10)$& \,\,\,$8.4^{\rm{o}} $  $\pm$  $ 1.4^{\rm{o}}$ \\
    \hline 
    \end{tabular}
 \tablecomments{For each parameter, the table shows both the prior 
 and posterior distributions resulting from the Bayesian inference. 
 $\mathcal{U}$ denotes uniform distribution. See Appendix.}
    \end{table*}

\begin{figure}
    \centering
\includegraphics[trim={0cm 0cm 0cm 0cm},clip,width=1.0\linewidth]{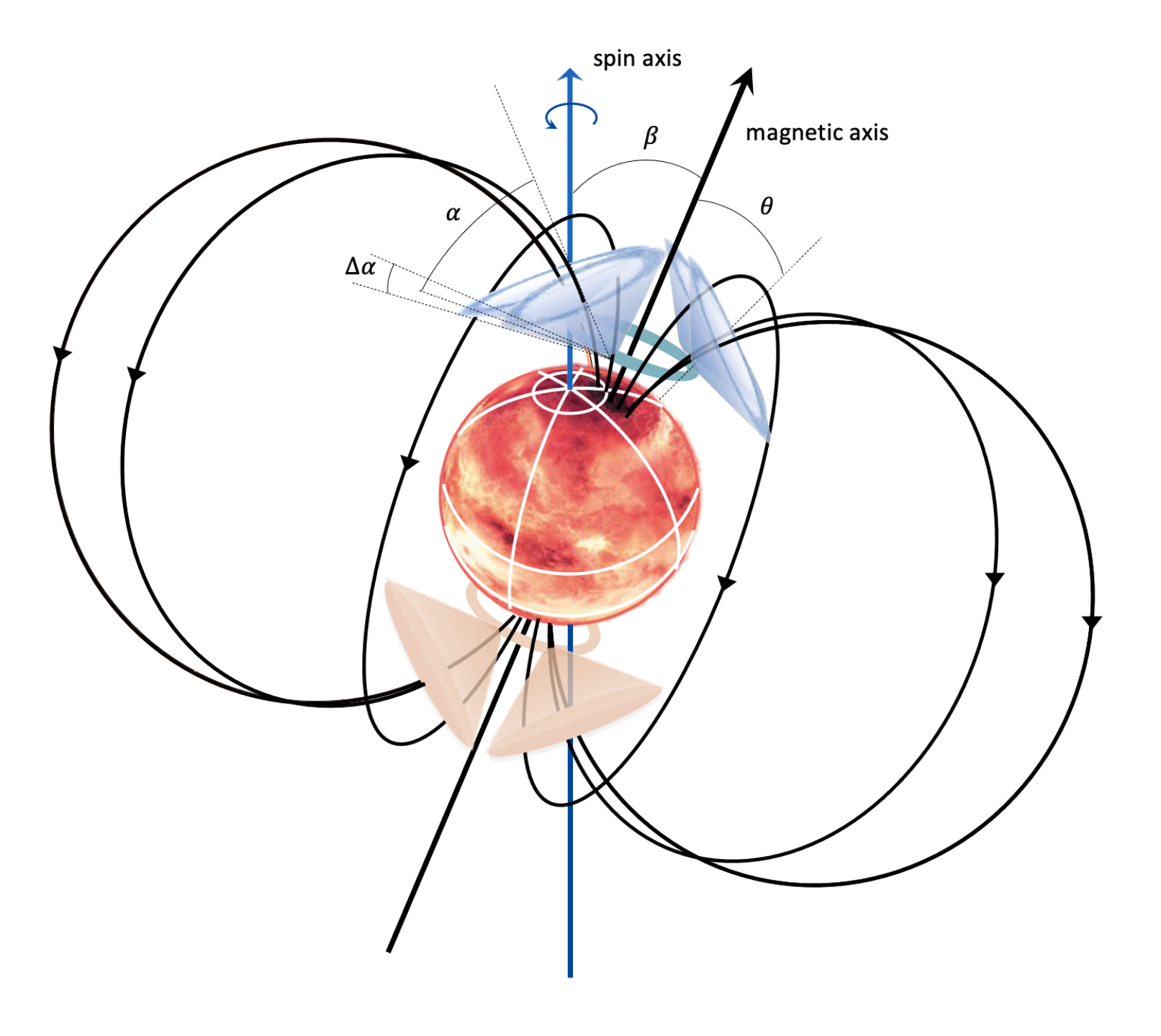}
    \caption{Representation of the auroral ring model for \J1122. The auroral rings are shown in both hemispheres; different colors indicates different handedness of the circular polarization (bluish/orange for RCP/LCP at the northern/southern magnetic hemisphere). Each point in the ring is an ECM emitter although, for clarity, only a few hollow-cone patterns are plotted. The values of magnetic axis obliquity, magnetic colatitude, and the geometry of the hollow cone correspond to the best fit model of Table \ref{table:mcmcresults}.  Notice that the spin axis is inclined towards the observer. The figure is intended for illustration purposes only, and it is not to scale. Image credit: Metazoa Studio/Hugo Salais (brown dwarf surface). Used with permission.}
\label{fig:illus}
\end{figure}

\subsection{The auroral ring model}

Several authors (\citealt{2012ApJ...760...59N}; \citet{2017MNRAS.470.4274T}) hypothesize that the quasi-steady ECM emission in UCDs, like that we detect in \J1122, follows a mechanism similar to that responsible for Jupiter auroral DAM emission 
(decameter non-Io emission; \citealt{2001P&SS...49.1067C}). Details of this model can be found in the references above; we outline them briefly here: the model assumes the existence of an equatorial plasma disc which corotates with the magnetosphere around the central object. The origin of the plasma is attributed to a companion object, with a role similar to that of Io in the Jovian system. The plasma expands radially outwards by centrifugal forces as its angular velocity decreases to conserve angular momentum. The shear in rotational flow resulting from the departure of the plasma from rigid corotation induces radially directed currents in the plasma sheet (associated with $E = -v\times B$, with $v$ the relative velocity between the plasma and the magnetosphere, and $B$ the local magnetic field). The presence of a gradient of the plasma angular velocity produces an azimuthal component of the magnetic field near the magnetic equator which effectively manifests in a bending of the magnetic field lines. The currents associated to this azimuthal magnetic field consist on upward-directed currents from the equator following the magnetic field lines towards higher magnetospheric latitudes, returning from the magnetosphere to the plasma sheet to close the circuit. The high-energy electrons in these upward field-aligned currents feed auroral phenomena. A similar circuit is induced in the southern magnetic hemisphere. This approach justifies Jupiter non-Io DAM auroras and is thought to be at work in UCDs.

\begin{figure*}[!t]
    \centering
\includegraphics[trim={0cm 2cm 0cm 2cm},clip, width=0.9\linewidth]{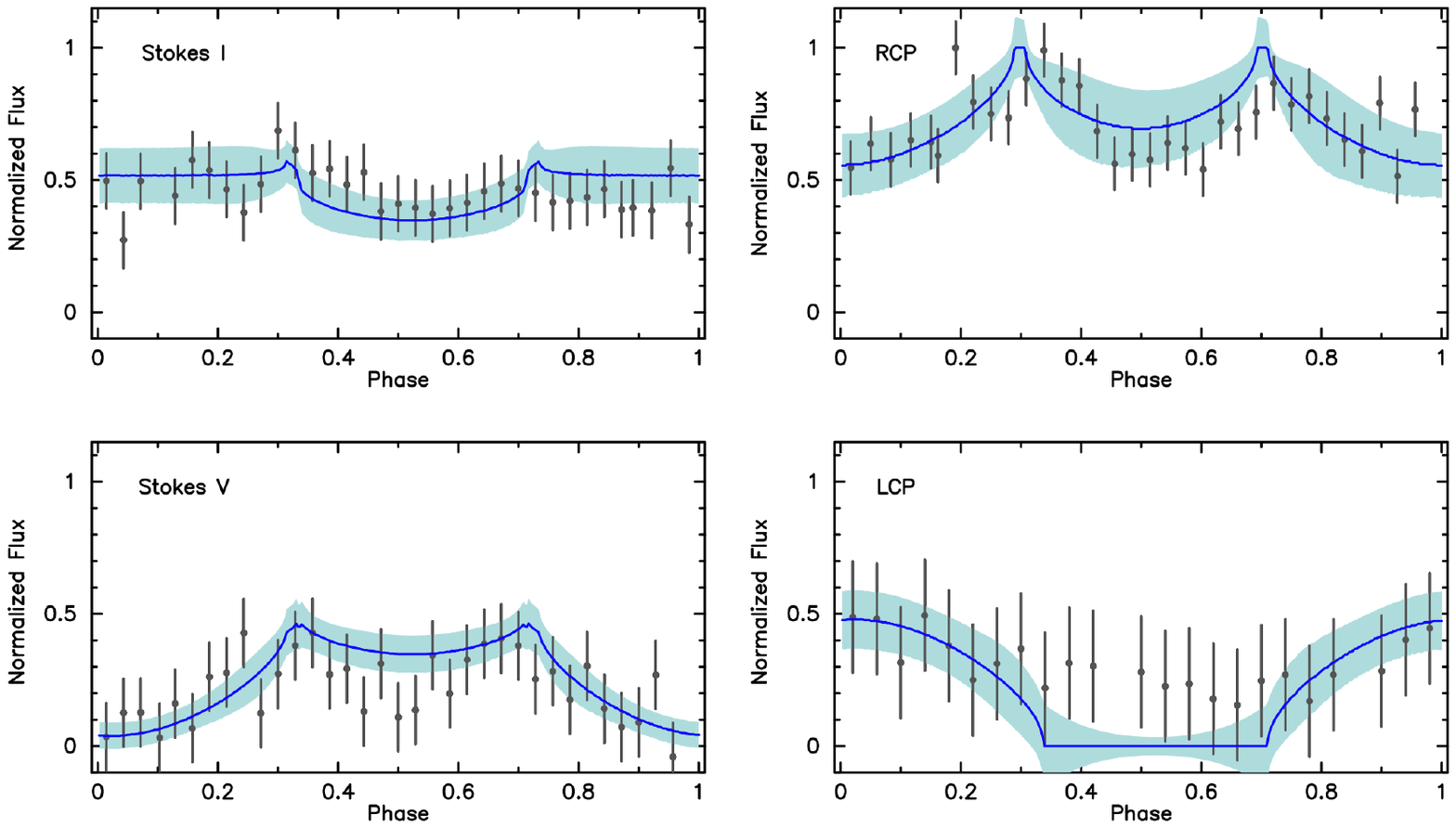}
    \caption{Normalized quasi-steady emission of \J1122. The data points corresponding the flare in segment C have been removed. For each panel, the lightcurve corresponding to the best fit resulting from our analysis is plotted in blue (see Table \ref{table:mcmcresults}). The shaded area shows $1\sigma$ variance of the model.}
\label{fig:modcurve}
\end{figure*}

Based on the physics described above, we followed the geometric model proposed in \citet{2024A&A...682A.170B} to reproduce the total and polarized lightcurves of \J1122. According to the references above, the auroral emission originates in circumpolar rings (which may remind the main oval in Jupiter) around the magnetic poles, where
the electrons supplied by the field-aligned currents bounce back to the equatorial plasma disk. The rings are loci of constant magnetic intensity that intersect the magnetic field lines at a certain magnetic colatitude. Every point belonging to the rings is assumed to be an emitter of ECM radiation along a conical shell whose axis is tangent to the local magnetic field line; all ECM emitters are considered equivalent in intensity and radiation pattern (i.e., same hollow cone). In a dipolar magnetic field, the radiation coming from the northern or southern auroral ovals will have different circular polarization, as $B$ would point in opposite directions with respect to the wave vector $\hat{k}$ of the emission ($B$ pointing outwards/inwards for the northern/southern magnetic hemisphere), hence producing a different handedness. The model assumes that $x$-mode emission dominates, meaning that, according to the IAU convention, the radiation from the northern/southern hemisphere is positive/negative Stokes V, i.e. RCP/LCP. The choice of the $o$-mode would have produced equivalent results, just with opposite handedness. 

The parameters involved in this geometrically oriented model are the inclination of the rotation axis ($i$), the obliquity of the magnetic field axis ($\beta$), the colatitude of the auroral oval ($\theta_B$), the half-opening angle of the ECM emission cone ($\alpha$), and the thickness of the conical shell ($\Delta\alpha$). For each point on the auroral rings, ECM radio emission is observed when the angle between the line of sight and the hollow cone axis is in the range $\alpha\pm\Delta\alpha$ (that is, the radio emission of the hollow cone reaches the observer). Observed radio emission from the northern/southern ring is counted as RCP/LCP, respectively. 

\begin{figure}
        \centering
      \includegraphics[trim={0cm 2cm 0cm 2cm}, clip, width=1.0\linewidth]{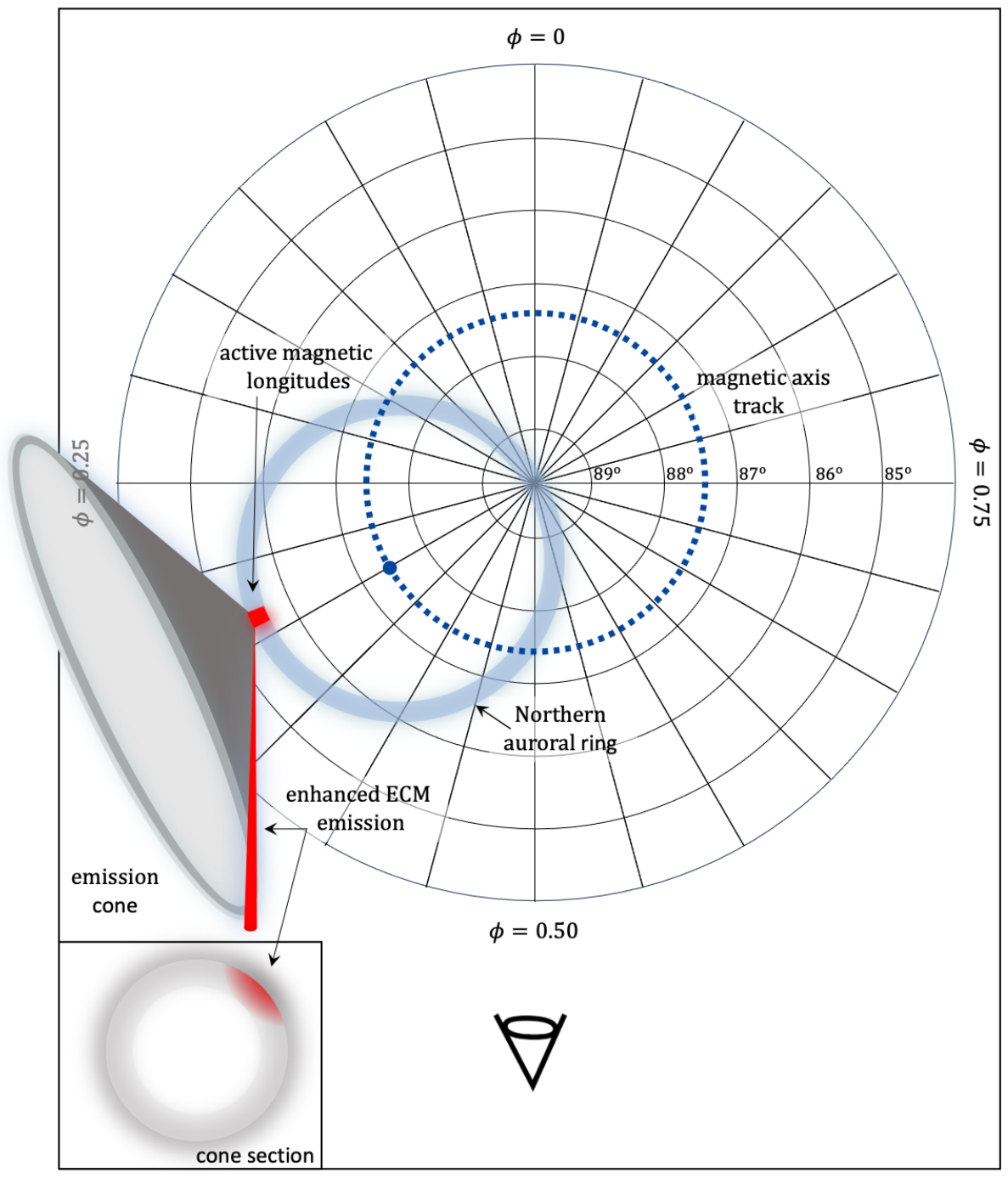}
    \caption{Polar projection of the northern aurora in \J1122. The grid represents brown-dwarf longitude and latitude (spin axis at latitude 90$\degr$). The track of the magnetic axis (blue dot) around the spin axis is plotted in dark-blue dashed line; the light-blue ring corresponds to the oblique projection of the northern auroral ring.  The active magnetic longitudes are marked in red color on the auroral ring; the enhanced ECM emission originating at these active longitudes (reddish color on the hollow cone) is visible by the observer, generating the large flare seen in segment C of our VLBA series (epoch 2022.82). The configuration corresponds to a rotation phase of $\sim$0.3; as the brown dwarfs rotates, the enhanced emission is not pointing to the observer anymore, which in turn suppresses a second flare within the same rotation.  Inset: section of the hollow cone showing the hot spot postulated to reproduce the lightcurve seen in epoch 2022.82. }
   \label{fig:grid}
\end{figure}
 
Taking both the Stokes I and RCP averaged lightcurves as input, we used the \texttt{magnetic-geometry-retrieval} code\footnote{\url{https://github.com/robkavanagh/magnetic-geometry-retrieval/}} \citep{2024A&A...692A..66K} to estimate the most probable values of the geometric parameters defined above (details are provided in the Appendix). The results of our analysis are shown in Table \ref{table:mcmcresults}, and illustrated in Fig. \ref{fig:illus}. The spin axis is slightly inclined toward the observer ($i=86.8^o$), with a magnetic axis obliquity of $\beta=2.6^o$, so that radiation from the northern magnetic hemisphere has a better chance to reach the observer, as expected by the predominance of RCP. Likewise, the weaker LCP flux densities seen in Fig. \ref{fig:lcurves} are a consequence of the less favorable orientation of the LCP radiation patterns originating at the southern auroral ring. Both RCP and LCP alternate during the rotation period, due to the changing orientation of the magnetic axis as the brown dwarf rotates. 

With respect to the beaming parameters, we find a half-opening angle of 
$\alpha=86.0^o$ and a cone thickness of $\Delta\alpha=8.4^o$. In comparison, the pattern of the Jupiter ECM radio emission is observed to have a similar half-opening angle 
($\alpha\sim70-90^o$) with a thinner conical sheet ($\Delta\alpha\sim1^o$; \citealt{2012P&SS...61...32C}). On the other hand, \citet{2024A&A...692A..66K} report 
$\alpha\sim70^o$ and $\Delta\alpha\sim4^o$ for the emission pattern of a T8-dwarf. 
Although our estimates are in reasonable agreement with the values above, the validity of these comparisons might be limited by the complex scenario in the maser region, with  different electron velocities and/or different directions of the magnetic field.

We used the values reported in Table \ref{table:mcmcresults} to model the observed 
lightcurves (see Fig. \ref{fig:modcurve}), whose main features at different polarizations are well reproduced. The peaks of the double lobe seen in RCP (and consequently seen in Stokes I and V) effectively correspond to the rotation phases ($\sim$0.3 and $\sim$0.7) which maximize the range of brown dwarf magnetic longitudes whose ECM emission reach the observer. The different slopes of the RCP and Stokes V lightcurve at the extremes of the rotation phase are due to the contribution of LCP flux density. For the same reason, the Stokes I curve is flatter than the RCP curve. The overall agreement of the proposed auroral ring model and our data is remarkable. 

\subsection{A closer look to the flare of 2022.82 }

The RCP lightcurve corresponding to segment C (epoch 2022.83) of our VLBA series shows a 100\% polarized burst with flux density $>$ 1\,mJy that peaks in rotation phase 0.3. The duty cycle of the flare is $\sim$30\% (larger than that of most UCDs; \citealt{2016ApJ...818...24K}) and it was observed during two consecutive rotations at the same phase with decreasing intensity (from $\sim$1100 to 700\,$\mu$Jy after one turn; see Fig. \ref{fig:Cflare}). 

The flare seen in \J1122 resembles similar phenomena seen in the Jupiter non-Io DAM emission, which also vary in time scales of hours. Indeed, non-Io auroras in Jupiter show a modulation in intensity that is attributed to the expansion or compression of the magnetosphere induced by the solar wind \citep{2003P&SS...51...57C}. In addition, periodic bursts have been observed at determined rotation phases \citep{2011pre7.conf..157P}, suggesting the existence of active Jovian magnetic longitudes, likely corresponding to areas of the plasma sheet in subcorotation with the magnetosphere with a more efficient generation of ECM emission. The flare seen in \J1122 seems to fit to this latter periodic type of event, considering that the flare occurs at a particular rotation phase, which points towards a preferred range of longitudes producing enhanced ECM emission. A potential drawback of this interpretation is that we would expect to observe the flare twice per
rotation (corresponding to the dawn and dusk of the ECM hollow cone emission pattern attached to the active longitudes), while only one event is observed in our lightcurve. Again, taking the non-Io emission as a guide, we find that some periodic non-Io bursts in Jupiter behave similarly, with only one peak appearing per rotation \citep{2013P&SS...77....3P}, being the second one absent or strongly weakened. This dawn-dusk side asymmetry of the burst suggests the presence of anisotropies in the hollow cone pattern of the enhanced ECM emission. Such anisotropies could be expected given the non-linear dependence of the maser growth with a large number of parameters (plasma density, frequency, velocity distribution, etc); therefore, small changes of some of the parameters above may produce a substantial variation in the maser intensity and/or directivity.

\begin{figure}
    \centering
\includegraphics[trim={1.0cm 2.0cm 8.5cm 19cm},clip, width=1.0\linewidth]{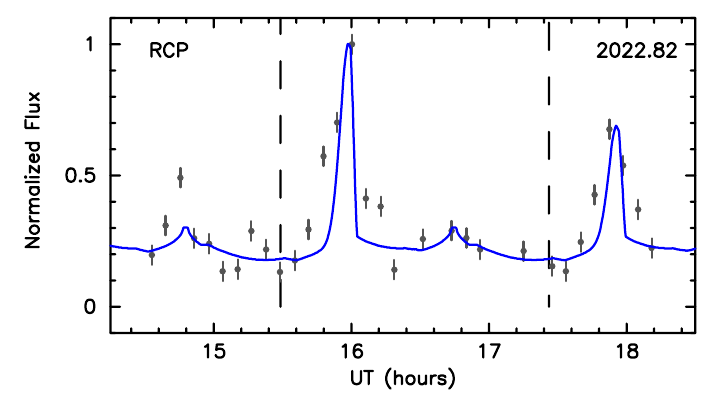}
    \caption{Following the construction shown in Fig. \ref{fig:grid}, we qualitatively reproduce the lightcurve of segment C (blue continuous line). The model consists of a quasi-steady, non-flaring, auroral component (according to our best fit parameters shown in Table \ref{table:mcmcresults}) plus a flaring radiation originating in a certain range of active longitudes (10\degr\, in size) of the northern auroral ring. The largest flare 
    at $\sim$16:00\,UT is 
    replicated assuming these active longitudes radiate following an anisotropic hollow cone pattern. The anisotropy consists on an enhancement ($\times$\,50 with respect the average emission) of a portion of the conical sheet (10\degr\, in size; see Fig. \ref{fig:grid}). The decrease of the flare intensity in the third period ($\sim$18:00\,UT) is reproduced by reducing the range of active longitudes to 5\degr. Multiple solutions are possible using a different range of active longitudes, burst intensities or burst sizes. This plot shows that a simple anisotropy may generate only one flare per rotation.} 
 \label{fig:tflare}
\end{figure}

Following this line of reasoning, we used the geometric model shown in the previous section to simulate the effect of an anisotropic pattern able to explain the pulse(s) profile in Fig. \ref{fig:Cflare} (i.e. absence of a second pulse in the same rotation). For simplicity, we simulated an increase of the intensity of the ECM emission in a range of magnetic longitudes; these active longitudes emit enhanced emission only in a particular direction (chosen to be visible by the observer), in practice generating an anisotropic hollow-cone pattern (see Fig. \ref{fig:grid} for details). The dusk/dawn asymmetry introduced by this basic anisotropy suffices to suppress (or weaken) one of the two pulses, reproducing the RCP lightcurve seen in 2022.82 (see Fig. \ref{fig:tflare}).  Other more elaborate anisotropies, whose application to \J1122 is beyond the scope of this paper, have been considered in the case of Io-related ECM emission. Galopeau \& Boudjada (2011) proposed that Io-related ECM bursty emission could be radiated in a flattened hollow cone, i.e., a hollow cone with elliptical section whose minor axis is aligned with the local magnetic field. Similarly, a {\it tangent plane beaming model} has been proposed for the terrestrial Auroral Kilometric Radiation \citep{2008GeoRL..35.7104M} and the magnetic chemically peculiar star CU Vir \citep{2011ApJ...739L..10T}. These patterns restrict the burst emission to a particular active rotation phase, resulting in a single peak per rotation in the lightcurve.

We notice that the burst seen in \J1122 reaches flux densities values ($\sim$mJy) similar to those reported by RW16, although some differences appear when comparing both events. First, the handedness of the flares reported in RW16 is LCP, opposite to that in segment C; second, the duration of the RW16 flares is 30--120\,s, very impulsive flares in comparison with the gradual behavior of the VLBA flare, with a duration $\sim$700\,s; and 
third, the detection rate in our VLBA campaign is substantially lower, while RW16 reported 5 flares in 29 observing hours, we only detected two pulses (considering two consecutive rotations in segment C) in 20 observing hours (although we may have missed other events, given our 6\,min long observing duty cycle). WGB17 did not report any 
mJy bursts, indicating that the occurrence of these bursts is very occasional. The differences stated above suggest that the VLBA RCP event reported in this paper and the Arecibo LCP flares in RW16 may respond to the action of different mechanisms releasing magnetic energy into the atmosphere of \J1122. However, a larger statistics of flares in this source would be needed to clarify their nature.

\begin{table*}[t]
\centering
    \caption{\J1122 J2000.0 coordinates} 
    \label{table:astropos} 
    \centering 
    \begin{tabular}{l l c c} 
    \hline\hline 
    Instrument & Program (epoch) & Right Ascension (h\,\,m\,\,s) & Declination ($\degr$\,\,$'$\,\,$"$)\\ 
    \hline 
     VLA$^{\mathrm{a}}$  &   16A-463 (2016.36) & 11\,\,22\,\,54.260\,\,$\pm$\,\,0.015 & 25\,\,50\,\,20.0\,\,$\pm$\,\,0.2\\
     VLBA &    BG278\,A (2022.27) & 11\,\,22\,\,53.819532\,\,$\pm$\,\,0.000007 & 25\,\,50\,\,18.09600\,\,$\pm$\,\,0.00013\\
     VLBA &    BG278\,B (2022.51) & 11\,\,22\,\,53.799128\,\,$\pm$\,\,0.000007 & 25\,\,50\,\,17.99740\,\,$\pm$\,\,0.00013\\
    VLBA &    BG278\,C (2022.82) & 11\,\,22\,\,53.783155\,\,$\pm$\,\,0.000007& 25\,\,50\,\,17.85634\,\,$\pm$\,\,0.00013\\
     VLBA &   BG278\,D (2023.02) & 11\,\,22\,\,53.768489\,\,$\pm$\,\,0.000007 & 25\,\,50\,\,17.81338\,\,$\pm$\,\,0.00015\\
     VLBA &   BG278\,E (2023.29) & 11\,\,22\,\,53.742875\,\,$\pm$\,\,0.000007 & 25\,\,50\,\,17.76898\,\,$\pm$\,\,0.00015\\

    \hline 
    \end{tabular}
    \tablecomments{$^{\mathrm{a}}$ \citet{2017ApJ...834..117W}.\\}
    \end{table*}

\subsection{Astrometric analysis}

As we explained in Sect. 2, the phase-reference mapping process conserved the positional information of \J1122 with respect to
the external quasar J112553.7$+$261019. Therefore, the brightness peak of the phase-referenced maps of \J1122 provides precise absolute coordinates referred to the ICRF. To optimize the astrometric precision, we used the RCP phase-referenced maps, the data set with the highest SNR (the use of the Stokes I maps provides compatible, but less precise positions). 
We performed an error analysis to determine the uncertainties of the radio coordinates due to errors inherent in the propagation media, the reference source structure, and the geometry of the interferometric array. We also considered the uncertainty associated with the location of the ECM emission originated in the auroral rings, in turn located at a determined height above the object surface. This height is determined by the gyrofrequency of the electrons; for M dwarfs (Bloot et al. 2024), heights for the fundamental frequency or its second harmonic may reach 1.2 stellar radii above the object surface. Assuming a similar scaling for \J1122 (height of 1.2\,R$_{\rm{J}}$), this translates to an uncertainty of 0.07\,mas at the distance of the brown dwarf.
All combined, these systematic contributions were five times larger than the limiting thermal noise uncertainty associated with the peak of brightness of the VLBI maps (see e.g. \citealt{2015A&A...578A..16A}). We determine a final astrometric error budget of $\sim$0.15 mas in each coordinate. The resulting positions of \J1122 and their corresponding standard deviations are shown in Table \ref{table:astropos}.

\begin{figure}[!ht]
    \centering
\includegraphics[trim={6cm 1.5cm 5cm 2.5cm},clip, width=1.0\linewidth]{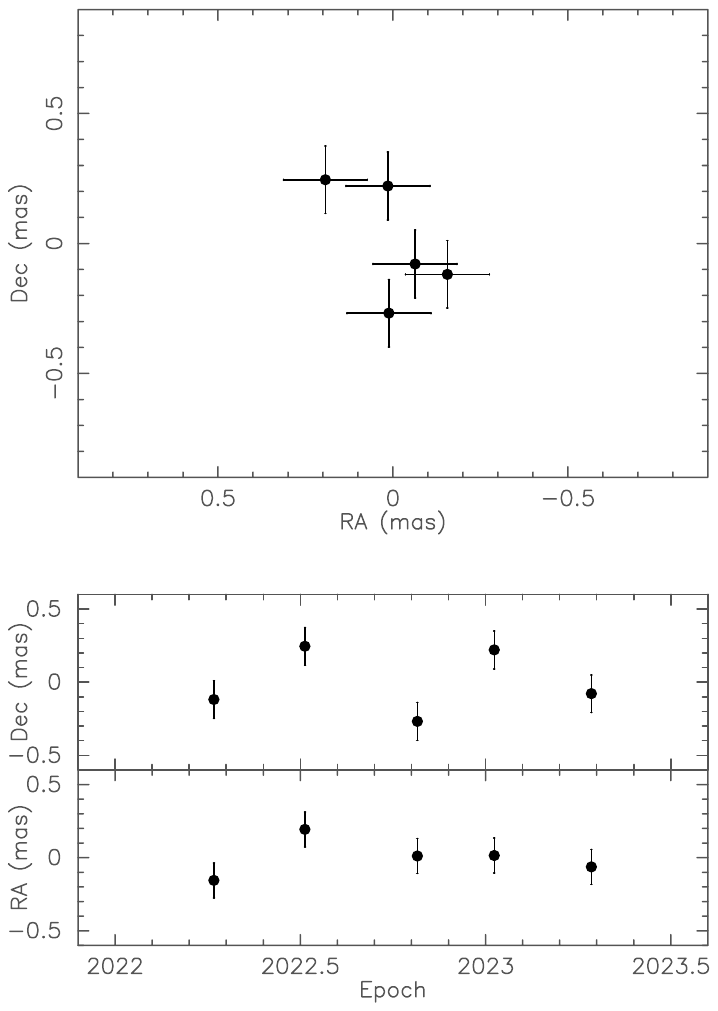}
    \caption{Astrometric residuals \J1122 in right ascension and declination after subtracting
proper-motion and parallax effects, shown as sky motion (upper panel) and time dependence (lower panel).}
\label{fig:astrores}
\end{figure}

The VLBA coordinates of \J1122 were used to derive
the proper motion, parallax via a weighted least-squares fit. We augmented our data set including  
the VLA position given in WGB17, despite of its relatively large 
uncertainty, to better constrain the large proper motion of this object. The derived astrometric parameters are shown in Table \ref{table:astrompar}, and the residual postfit positions after removing the proper motion and parallax effects in Fig. \ref{fig:astrores}. Our estimates agree within uncertainties with those provided 
by \citet{2020AJ....159..257B}, providing an improvement in precision of one order of magnitude.

\subsubsection{The wide companion LHS\,302}

\J1122 and the M5 dwarf LHS\,302 exhibit similar proper motion and parallax values; they constitute a wide pair  with a projected separation of $\sim$4500\,AU (265 arcsec; \citealt{2011ApJS..197...19K}). The proper motion and parallax of both objects have been subsequently improved \citep{2020AJ....159..257B} validating their similar kinematics. Our astrometric analysis of \J1122, along with the Gaia DR3 astrometric parameters available for LHS\,302, enables a further, submilliarcsecond precise comparison 
of the kinematics of both objects. The values shown in Table \ref{table:astrompar} confirm a common distance (within uncertainties) and a very similar proper motion, with differences smaller than 0.5\%. This coincidence supports the hypothesis that \J1122 and LHS\,302 belong to the same moving group (loose associations of coeval, co-moving stars; \citealt{2004ARA&A..42..685Z}). In addition, the similar kinematics of this co-moving pair of M-dwarf and brown dwarf may provide clues of the formation mechanism. \citet{2007A&A...466..943G} propose that brown dwarfs may form as distant companions to M-dwarfs by disc fragmentation; however, at large separations ($>$100\,AU), both objects are not strongly bounded and the brown dwarf may be easily disrupted by close encounters with passing stars, finally producing a co-moving wide, low-mass binary system, just as LHS\,302 and \J1122. The remarkable kinematical coincidence of both objects seems to support this hypothesis. We notice that, in this {\it gentle disruption} scenario (following \citealt{2007A&A...466..943G}), \J1122 may retain a circumstellar disc, therefore, the presence of companions to the brown dwarf is not excluded. We discuss this in turn. 

\subsubsection{Bounds to the presence of companions to \J1122}

\begin{figure}
    \centering
\includegraphics[trim={0cm 4.0cm 0cm 4.0cm},clip,width=1.0\linewidth]{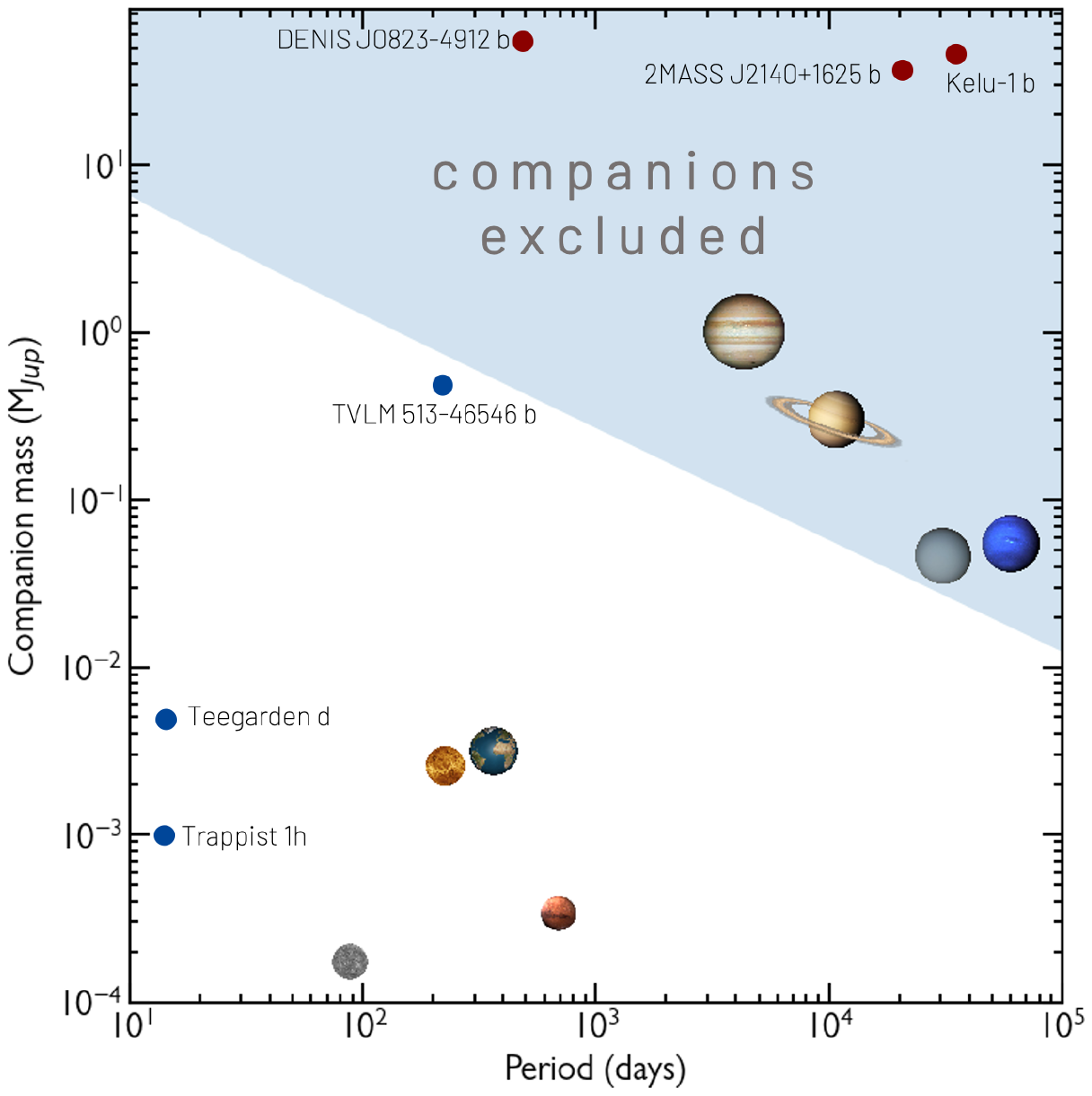}
    \caption{Companions excluded by our astrometric analysis. The shaded area shows the space of companion masses and periods producing a reflex motion larger than the rms of the astrometric residuals ($\sim$0.3\,mas). Red dots indicate a selection of known binary brown dwarfs; blue dots indicate a selection of brown dwarfs with known exoplanets; solar system planets are shown for reference. Our astrometric fit discards \J1122 to be a binary brown dwarf, but it is largely compatible with the presence of an Earth-to-Saturn mass exoplanet.}
    \label{fig:maper}
\end{figure}

Proper motion and parallax suffice to a great extent to account for the trajectory of \J1122 on the sky. However, the root-mean-square (rms) of the postfit residuals in Fig. \ref{fig:astrores} is $0.3$\,mas, comparable but larger than the uncertainties of the coordinates ($\sim2\sigma$), 
thus indicating possible submilliacsecond departures from the linear motion which might be the consequence of a gravitational reflex motion. While an orbital fitting requires a larger data set, the dispersion of our astrometric residuals imposes immediate bounds to the presence of low-mass companions around \J1122 through Kepler's Third Law expressed as 
$a_{\ast}=\pi\,m_{c} ( P^2/(m_{\ast} + m_{c}))^{2/3} $, with a$_{\ast}$ the amplitude of 
the reflex motion of \J1122, $\pi$ the parallax, $P$ the period of the orbit (yr), and $m_{\ast}$ and $m_{c}$ the masses of \J1122 and the companion (M$_{\odot}$), respectively.\\
Assuming face-on, circular orbits, companions with mass $m_{c}$ and orbital period $P$ 
producing a reflex motion with amplitude $a_{\ast}$ larger than the scatter of our astrometric residuals (0.3\,mas) would be excluded; this population corresponds to the shaded area of Fig. \ref{fig:maper}. Our data seem to discard companions with large values of the mass ratio $q$ ($q>0.1$; unless periods of a few days are considered). As an example, giant planets similar to those of the solar system are excluded, 
but the presence of hot Jupiters and Earth-like planets is largely allowed 
according to Fig. \ref{fig:maper} (in particular, a rocky planet, that might constitute a potential source of plasma in the scenario discussed in Sect. 4.2). Saturn-like planets, similar to that discovered around TVLM\,513$-$46546 \citep{2020AJ....160...97C}, would straddle the border of the exclusion area. As \J1122 shows as an excellent astrometric target, future observing campaigns will surely refine the constraints provided in this paper.\\

\begin{table}[t]
\caption{Astrometric parameters of \J1122 and its wide companion LHS\,302}   
\label{table:astrompar} 
    \centering 
    \begin{tabular}{lcc} 
    \hline\hline 
& \J1122$^{(a)}$ & LHS\,302$^{(b)}$ \\
\hline
$\mu_{\alpha}$ (mas\,yr$^{-1}$) & -1015.62 $\pm$   0.14 & -1010.9189  $\pm$  0.0734  \\
$\mu_{\delta}$ (mas\,yr$^{-1}$) & \,\,\,-322.08 $\pm$   0.20 & \,\,\,-323.1270  $\pm$  0.0684  \\
Parallax (mas) & \,\,\,\,\,\,\,\,61.68  $\pm$  0.10 & \,\,\,\,\,\,\,\,61.6520  $\pm$   0.0644 \\
    \hline 
    \end{tabular}
    \tablecomments{$^{(a)}$ This paper. $^{(b)}$ Gaia DR3 archive. }
    \end{table}
      
\section{Conclusions}\label{sect:conclusion}

A series of 5\,GHz VLBA phase-referenced observations of the T6 brown dwarf \J1122 shows that this very cool object is a persistent radio emitter, detected at all observing epochs, with radio emission originating in a very compact region, bright enough to be detected with VLBI. Our main findings can be summarized as follows:

\begin{enumerate}
    \item The milliarcsecond-scale maps of \J1122 consist of a single unresolved component with circularly polarized radio emission. This radio emission is predominantly RCP, although traces of compact LCP emission are detected in some epochs. 
    \item We examined the time series of the RCP data set for variability and found a preferred periodicity of 1.95$\pm$0.03 hr, which we assume corresponds to the rotation period of \J1122. The main feature of the combined, averaged lightcurves, phase-wrapped to the rotation period, is a double-peak feature in RCP; the handedness of the polarization seems to alternate throughout the rotation period, with a weak increase of the LCP flux density in coincidence with the depths in the RCP lightcurve. We interpret these lightcurves as the quasi-steady auroral radio emission of \J1122. 
    \item The combined total and polarized lightcurves of \J1122 can be reproduced with a geometric model which assumes that the ECM emission is produced in circumpolar auroral rings, similar to the main oval auroras in Jupiter. The model fits remarkably well the lightcurves providing estimates of the spin axis inclination, the tilt of the magnetic axis and the geometry of the ECM radiation pattern. If present, other radiation mechanisms apart of ECM have a residuals role.
    \item A large 100\% RCP flare is detected in our VLBA epoch 2022.82. The flare is seen only once per rotation, which, according to the geometric parameters derived by us, may imply the existence of anisotropies in the ECM radiation pattern.
    \item Our VLBA phase-referencing analysis provides submilliarcsecond-precise revised values of proper motion and parallax of \J1122. We have confirmed the common kinematics of \J1122 and the M-dwarf LHS\,302. According to theories of brown dwarf formation, this co-moving pair of objects may have formed together, being the brown dwarf gently disrupted by gravitational interaction with passing stars. 
    \item The astrometric residuals in right ascension and declination impose a limit of 0.3\,mas for the reflex motion of \J1122 due to a possible, unseen low-mass companion. Unless very short periods are considered, this limit effectively excludes companions larger than Saturn around \J1122.\\
    
\end{enumerate}

The results obtained for \J1122 emphasize the similarity between the magnetic phenomena occurring in substellar objects and those 
occurring in the planets of our solar system, Jupiter in particular. Indeed, very recently, radiation belts have been detected around 
the ultracool dwarf LSR J1835+3259 (\citealt{2023Natur.619..272K}, \citealt{2023Sci...381.1120C}), a scaled-up version of the Jovian belts. And, in this paper we demonstrate the 
presence of persistent main-oval auroras in \J1122, which may respond to a very similar mechanism as those known in Jupiter. 
Both results support the analogy between radio emitting brown dwarfs and the Jupiter system. Such analogy would be complete with a confirmed detection of magnetic star-planet interaction similar to the Jovian Io auroras. 

\begin{acknowledgements}
The authors thank the anonymous referee for the thorough, and constructive review of this manuscript, which improved the overall quality of the article. This work is based on observations with the Very Long Baseline Array (VLBA), which is operated by the National Radio Astronomy
Observatory (NRAO). The NRAO is a facility of the National
Science Foundation operated under cooperative agreement by
Associated Universities, Inc. Scientific results from data presented in this publication are derived from the VLBA project code BG278. This work has
made use of data from the European Space Agency (ESA) mission
Gaia, processed by the Gaia Data Processing and Analysis
Consortium (DPAC). Funding for the DPAC has been provided
by national institutions, in particular the institutions participating
in the Gaia Multilateral Agreement. JCG, JBM, and JMM were supported
by projects PID2020-117404GB-C22 and PID2023-147883NB-C22 funded by MCIN/AEI,
CIPROM/2022/64, funded by the Generalitat Valenciana, and
by the Astrophysics and High Energy Physics programme by
MCIN, with funding from European Union NextGenerationEU
(PRTR-C17.I1) and the Generalitat Valenciana through grant ASFAE/2022/018.
MPT and LPM were supported by projects PID2020-117404GB-C21, PID2023-147883NB-C21, and CEX2021-001131-S funded by MCIN/AEI and by the European Union
(NextGenerationEU grant PRTR-C17.I1). This research made use of Astropy, 
a community-developed core Python package for astronomy \citep{2018AJ....156..123A}.
\end{acknowledgements}

\appendix
\section{Estimate of the geometric parameters of the auroral ring model}
\begin{figure*}[h]
    \centering
\includegraphics[trim={0cm 0.5cm 0cm 0.0cm},clip,width=0.8\linewidth]{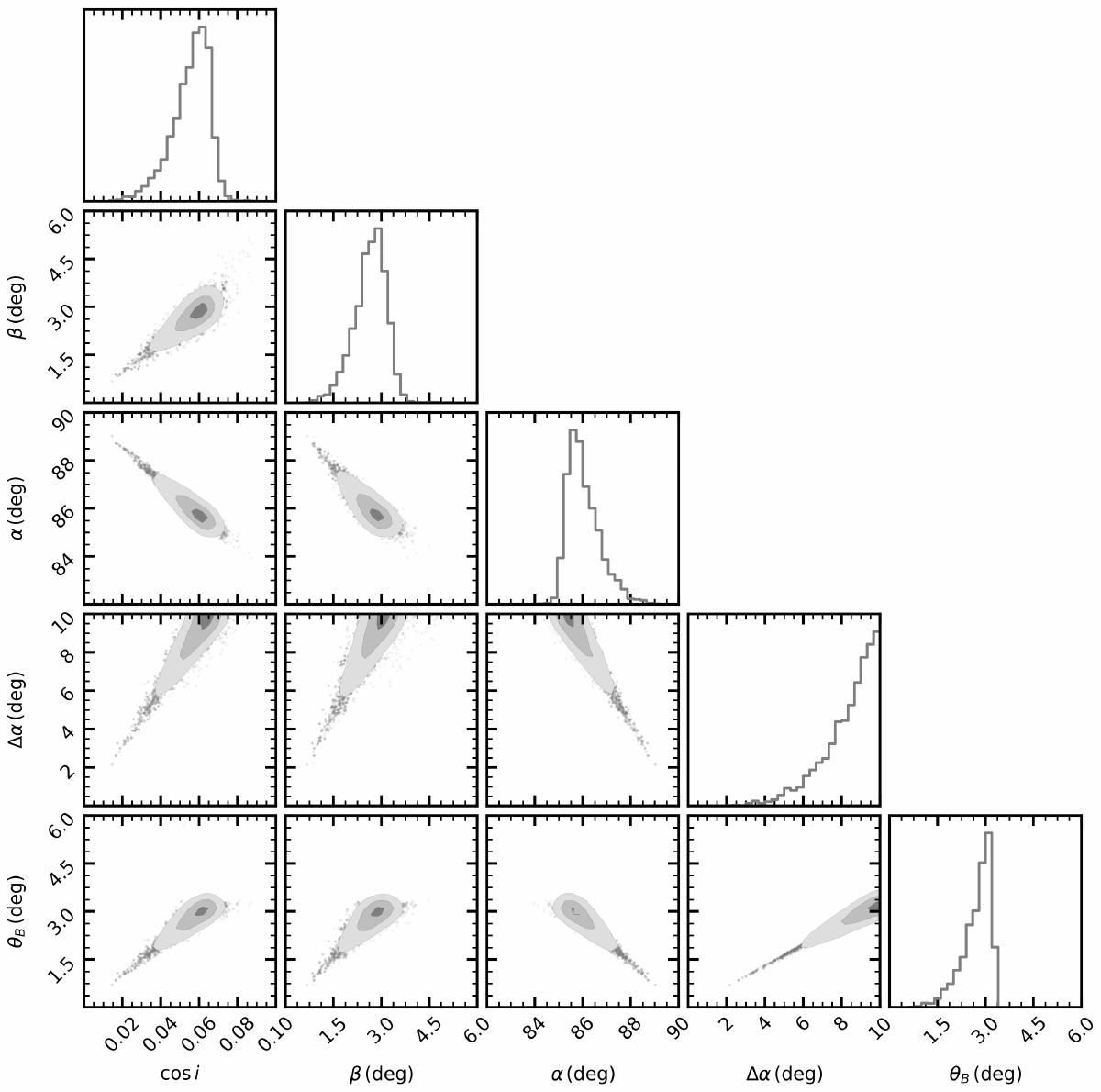}

\caption{Corner plot showing the results of our parameter estimation for the auroral ring model. The histograms on the diagonal show the posterior distribution for each parameter. Off-diagonal plots are two-dimensional projections of the sample showing covariances; darker shading denotes regions of higher probability densities.}
    \label{fig:cornerplot}
\end{figure*}

In Sect. 4.2, we used the \texttt{magnetic-geometry-retrieval} code\footnote{\url{https://github.com/robkavanagh/magnetic-geometry-retrieval/}} \citep{2024A&A...692A..66K} to derive the most probable parameters corresponding to the orientation of \J1122, and the geometry of the ECM hollow-cone pattern. This code uses Ultranest\footnote{https://johannesbuchner.github.io/UltraNest/}, a Monte Carlo-based algorithm which efficiently samples a wide range of the parameter space and provides posterior distributions for each model parameter via Bayesian inference. The code assumes a dipolar magnetic field corotating with the brown dwarf where the radio emission originates in a number of active field lines. This feature was not used in the case of \J1122, as we consider the auroral rings to be uniform radio emitters, in the line described in \citet{2024A&A...682A.170B}. We also simplified the code to work with a single frequency band. The averaged Stokes I and RCP data were jointly used as input. We uniformly sampled the following range of the parametric space: 
$0\degr \leq i\leq 90\degr$; $0\degr \leq \beta \leq 60\degr$; $30\degr \leq \alpha \leq 90\degr$; $0\leq \Delta\alpha \leq 10\degr$; and $0\leq \theta_B \leq 40\degr$. The prior in $\Delta\alpha$ was imposed to find solutions with relatively small cone thickness (given the values reported for Jupiter and other T-dwarfs; see Sect. 4.2). 
With the constraints above, we found convergence in 75 steps. The best-fit parameter estimates are shown in Table \ref{table:mcmcresults}, which provide a reduced $\chi^2$ of 1.31. The corresponding posterior distributions are shown in Fig. \ref{fig:cornerplot}. As a result of the prior imposed for $\Delta\alpha$, we notice that the posterior distribution of some of the fitted parameters is asymmetric. This may reflect the limited sensitivity of our averaged data to small cone thicknesses. Despite this limitation, the comparison between observed and modeled lightcurves is satisfactory (see Fig. \ref{fig:modcurve}).

\bibliography{manuscript}{}
\bibliographystyle{aasjournal}

\end{document}